# Scalability in Workforce Management: Applying Scalability Principles to Foster a Four-Day Work Week

Sunkanmi Oluwadare

University of Cincinnati, oluwadse@mail.uc.edu

Ebubechukwu Edokwe

University of Cincinnati, edokweeu@mail.uc.edu

Olatunde Ayeomoni

University of Cincinnati, ayeomooe@mail.uc.edu

## ABSTRACT

The traditional five-day workweek faces mounting challenges, prompting exploration of alternative models like the four-day workweek. This research explores the transformative potential of scalability principles derived from cloud computing and IT in redefining workforce management for a four-day workweek. The study employs a Multivocal Literacy Research methodology, combining grey literature and systematic review approaches. Through a comprehensive review of related work, the challenges, and benefits of transitioning to a four-day workweek are explored. Pilot programs, clear communication, and agility are identified as critical success factors. The synthesis of scalability principles in workforce management serves as a powerful framework for a smooth transition towards a four-day workweek. By prioritizing adaptability, dynamic resource allocation, and data-driven insights, organizations can unlock the full potential of a compressed work schedule. This research contributes valuable insights for organizations seeking to thrive in the evolving landscape of modern work structures and prioritizing employee well-being.

## 1  INTRODUCTION

Scalability refers to a cloud layer's capability to expand its usage of lower-layer services, thereby increasing its capacity [15]. Scalability, as witnessed in cloud computing, revolves around the dynamic allocation and de-allocation of resources based on demand [1]. It reduces manual involvement of provisioning servers or resources when there is a need to have one. There are 2 types of scalabilities in the cloud the Horizontal scalability which involves adding more servers to handle the same task and Vertical scalability which involves increasing the capacity of a single server by adding more resources [1]. Building upon these foundational concepts, the core principles of scalability include elasticity, flexibility, and resource optimization which emerge as critical focal points in system design [15].

There are factors that helps scalability to be optimal in the cloud computing, they include, implementing a notification system to alert administrators when adding or removing constraints can help in managing scalability effectively, Load balancing, Compatibility checks and Scalability according to user needs. [8].

Flexibility refers to the ability of a system to adapt to different types of workloads and changing requirements [15]. Resource optimization refers to the efficient use of resources to achieve maximum performance and cost-effectiveness. [15]. It is important to design systems that are scalable, by carefully considering factors such as elasticity, flexibility, and resource optimization. By doing so, cloud providers can ensure that their services can handle increasing workloads and changing requirements,

Translating this concept into workforce management involves assigning tasks in a manner similar to cloud resource allocation. This concept is referred to as "Workforce Scalability", it refers to the capacity of an organization to keep its human resources aligned by constantly transitioning from one human resource configuration to another in a seamless way [6]. This means that organizations can adjust their workforce size, competence mix, deployment pattern, and employee contributions as needed to stay competitive in dynamic situations.

In addition to the concept of workforce scalability, the critical significance of workforce agility emerges, especially for organizations navigating unpredictable changes. Workforce agility essentially empowers organizations to respond to market fluctuations swiftly and effectively, whether stemming from evolving demands, emerging product categories, or other external and dynamic factors [18]. Aligning with this notion, [12] shows the significance of identifying and deploying specific micro-tasks promptly, offering a crucial strategy for organizations to respond effectively not only to crises but also to the demands of the 'new normal.' This shows the critical need for businesses to not only possess agile workforces but also ensure their scalability, enabling them to adeptly maneuver through dynamic market landscapes."

Furthermore, workforce scalability is an important competency of an agile organization. It enables organizations to align and reconfigure their human resources on an efficient basis. The human resource configurations comprise four dimensions: competence mix, headcount, employee contribution, and deployment pattern [13].

The traditional work system, which prioritizes economic output over environmental concerns, has led to longer working hours, unemployment, and unsustainable practices that harm the environment and contribute to climate change [3]. The traditional 5-day 40-hour work week has several limitations and drawbacks. These include a negative impact on employee health and well-being, work-life balance, reduced productivity, and high turnover rates [5]. Long work hours and lack of flexibility can lead to stress, burnout, and other health problems, as well as conflicts between work and personal life [5]. Modern workplaces, including healthcare, are in the midst of a paradigm shift, exploring the viability of a four-day work week. [11]

Drawing inspiration from the scalability principles inherent in Information Technology (IT), traditionally associated with cloud computing, this research seeks to understand how scalable workforce management can serve as a catalyst for embracing a more flexible work schedule across diverse industries. Scalability in cloud computing allows organizations to dynamically adjust resources, ensuring efficient handling of varying workloads. This concept enables IT organizations to scale resources up or down based on industry needs, making the infrastructure more flexible and cost-effective.

In this context, the research question arises: How can organizations effectively leverage a four-day workweek by adopting scalability principles? By proposing a four-day workweek, the research aims to explore how organizations can effectively leverage scalability principles to optimize workforce management. This approach aligns with the scalability principles observed in cloud computing, where adaptability to changing demands, resource optimization, and enhanced overall system performance are key outcomes. Addressing scalability challenges and implementing effective solutions in workforce management could mirror the success seen in scalable and efficient cloud infrastructures.

## 1.1 Related Work

In our quest to identify related work, we conducted a targeted search across 2 databases, centering on articles related to the four-day work week. This section offers a concise overview, beginning with recent studies on the effects of the traditional five-day work week and concluding with insights from related work on the four-day work week.

It is a known fact that the five-day work week can have negative impacts on employees' health, well-being, work-life balance, productivity, and high turnover rates [5]. Additionally, [17] contribute to our understanding by highlighting several issues associated with the traditional five-day work week, including increased work-related stress, reduced social interaction among team members, and a low willingness to participate in meetings with social context due to the compressed working schedule. This reinforces the growing body of evidence pointing to the challenges posed by the conventional workweek and shows the need for exploring alternative structures like the four-day work week. The notion of the "five-day week" is fundamentally a social construct, offering no inherent guarantee of productivity. It simply mandates being in the office or at the desk until Friday at 5. This construct provides no insights into one's activities, time utilization, or the efficiency of work throughout the week [4].

Contrary to the prevailing notion that increasing work hours is a solution to productivity challenges, the work by [3] challenges this perspective by asserting that the most productive and affluent nations are those embracing shorter workweeks. This counters the current trend, highlighting that increasing productivity should, in fact, lead to the liberation of time for workers, a historical sign of societal progress and improvement rather than perpetuating the conventional and potentially detrimental five-day workweek.

As organizations seek innovative approaches to enhance employee satisfaction and productivity, the four-day work week has emerged as a promising paradigm shift. The transition to a compressed work schedule offers potential solutions to the challenges posed by the conventional work week [19].

Building on this momentum, recent studies show the interest in implementing a four day work week, In June 2022, 61 UK companies piloted a four-day workweek, with 92% continuing tests and 29% adopting it permanently by February 2023. The Harvard Business Review found a 1.4% rise in average revenue, a 57% decrease in employee turnover likelihood, and a 65% reduction in sick leave. This study offers compelling evidence for the positive impact of a four-day workweek [10]. Bersin [4] suggests that work redesign in the next big thing as many young workers do not abide or are not in support of the traditional work norms evident from the social media memes.

In exploring the implementation of a four-day workweek, a notable case study arises from the manufacturing industry, Advanced RV which participated in a global trial led by 4 Day Week Global, implementing a four-day workweek with maintained pay for 6 months. Despite initial skepticism, the transition increased employee satisfaction, fostering a healthier work-life balance. Efficiency improvements and the CEO's adaptability showcased the sector's openness to alternative work structures. After a year and a half, Advanced RV reported near recovery of productivity losses, highlighting sustained positive impacts. [9]

Alfares [2] examines the challenges and considerations involved in implementing a four-day workweek schedule, allowing employees to have either two or three consecutive days off per week. The research presents an optimization algorithm designed to determine the minimum workforce size and effectively assign employees to various days-off work patterns to minimize the total number or cost of workers allocated. Moreover, the study highlights the benefits of alternative work schedules, particularly emphasizing the advantages of the four-day workweek.

In examining the landscape of the four-day workweek, challenges and drawbacks emerge from historical attempts to compress a standard workweek into four days. Past initiatives witnessed fatigue, productivity declines, and operational difficulties, shedding light on the complex dynamics involved in transitioning to a condensed schedule [10]. Navigating the challenges of the non-linear work hours-productivity link and the need for substantial work redesign, organizations must streamline operations, redefine priorities, and embrace asynchronous communication. Resetting employee expectations is pivotal, involving clear communication, pilot programs, and productivity coaching [10].

The challenges of the four-day workweek emerge from studies showing that it may not universally enhance employee well-being. Issues include the potential for burnout, especially when compressing work into fewer days, and varying impacts depending on job types and work settings. Notably, Gallup's survey found that those on a four-day schedule did not report significantly higher well-being and reported higher burnout levels compared to those working five days. The effectiveness of a shorter workweek depends on factors such as individual preferences, job requirements, and the flexibility inherent in different roles [16].

As the article progresses, we propose to delve into potential solutions and adopt a transformative strategy for the implementation of a four-day workweek. This exploration aims to provide valuable insights into addressing challenges and capitalizing on the advantages of an alternative work structure, drawing parallels with the concept of scalability in the cloud.

## 2 METHODOLOGY

This study utilized the Multivocal Literacy Research methodology (MLR) proposed by [7] which is a combination of grey literature and the systematic literature review proposed by [14]. By employing MLR this study aims to capture the diverse perspectives and insights necessary to propose a comprehensive and inclusive approach to addressing workforce scalability and implementing a 4-day work schedule within organizations.

### 2.1 Planning

The preliminary stage of planning for the Multivocal Literature Review involves reading, understanding the scope and shaping the research question to meet the central aim of this review: comprehending how organizations can efficiently utilize a 4-day work week through the integration of scalability principles. After understanding the topic and shaping the research question, we started to answer the research question by first gathering literatures with the search query 'A four day work week' was applied for the formal search on Google Scholar, examining results of the first page. For the exploration of grey literature, the regular Google search engine was used, analyzing the results of the first page as well. This consistent approach allowed for a comparative assessment across both platforms.

"A Four Day Work Week"
Figure 1: Study's Search String

The study utilized Google Scholar for formal search and the regular Google search engine for accessing grey literature. Google Scholar, known for its academic and scholarly content, provided insights from academic papers. Meanwhile, the regular Google search engine broadened the scope, encompassing various sources beyond academic publications."

Table 1: Study Inclusion and Exclusion Criteria

| Number | Inclusion | Exclusion Criteria |
|---|---|---|
| 1 | Articles that describe the Four Day Work Week | Articles that are not publicly available. |
| 2 | Articles that appeared on the first page of the search engines | Articles that do not directly address the need for a four-day work week. |

## 2.2 Conducting the Review

The search query was employed to search the Google Scholar's database on December 12, 2023. The search yielded a total of 21 results. During the initial phase, 21 results were identified for the study, 10 For the formal search and 11 for the grey literatures. For the second phase, the exclusion criteria in Table 1 were applied by reading the titles and abstracts which resulted in 11 articles selected for the primary study. 4 Articles from grey literature were excluded because they didn't address the four-day work week and 6 were excluded for the formal search because they were not publicly available.

## 2.3 Study Selection

The primary studies for this research comprise the remaining 11 articles selected detailed in Table 2.

Table 2: Primary Study

| Title | Citation |
|---|---|
| Four-day workweek scheduling with two or three consecutive days off | Alfares, H. K. (2003). Four-day workweek scheduling with two or three consecutive days off. Journal of Mathematical Modelling and Algorithms, 2, 67-80. |
| A four-day workweek: a policy for improving employment and environmental conditions in Europe. | Ashford, N., & Kallis, G. (2013). A four-day workweek: a policy for improving employment and environmental conditions in Europe, The European Financial Review, 30 April 2013. |
| The Four-Day Work Week: An Idea Whose Time Has Come. | Bersin, J. (2023, November). The Four-Day Work Week: An Idea Whose Time Has Come. Retrieved from https://joshbersin.com/2023/11/the-four-day-work-week-an-idea-whose-time-has-come/ |

| Title | Citation |
|---|---|
| The four-day work week: a chronological, systematic review of the academic literature. Management Review Quarterly | Campbell, T. T. (2023). The four-day work week: a chronological, systematic review of the academic literature. Management Review Quarterly, 1-17. |
| A manufacturer tried the 4-day workweek for 5 days' pay and won't go back. | Hsu, A. (2023, November 11). Business: A manufacturer tried the 4-day workweek for 5 days' pay and won't go back. NPR. https://www.npr.org/2023/11/11/1207991399/4-four-day-work-week-manufacturing-work-life-balance |
| How to Actually Execute a 4-Day Workweek | https://hbr.org/2023/12/how-to-actually-execute-a-4-day-workweek |
| The 4-Day Work Week, Explained | https://www.gartner.com/en/articles/the-4-day-work-week-explained |
| Rapid-workforce-transformation-a-practical-guide-for-the-covid-19-crisis. | https://www.mckinsey.com/business-functions/organization/our-insights/rapid-workforce-transformation-a-practical-guide-for-the-covid-19-crisis. |
| Most people would love a four-day workweek. But it doesn't work for everyone. | Sahadi, J. (2023, November 7). Most people would love a four-day workweek. But it doesn't work for everyone. CNN. https://www.cnn.com/2023/11/07/success/four-day-workweek-survey/index.html. |
| How a 4-day work week and remote work affect agile software development teams. | Topp, J., Hille, J. H., Neumann, M., & Mötefindt, D. (2022, January). How a 4-day work week and remote work affect agile software development teams. In International Conference on Lean and Agile Software Development (pp. 61-77). Cham: Springer International Publishing. |
| Surprising Benefits of a Four-Day Week. | World Economic Forum. (2023). Surprising Benefits of a Four-Day Week. Retrieved from https://www.weforum.org/agenda/2023/10/surprising-benefits-four-day-week/ |

## 2.4 Quality Assessment

The quality assessment of the primary studies involved a thorough examination of the full articles to determine their relevance to our study's objectives. We developed criteria and checklists for assessing quality guided by our research goals, focusing on how well the studies aligned with our research questions. The comprehensive evaluation results, including assessment questions and outcomes, are presented in Table 3.

Table 3: Quality Assurance Criteria

| Quality Assurance Question | 1= Less Focus | 2= Some Focus | 3 = Most Focus |
|---|---|---|---|
| To what degree does the article focus on the Four-day work week | 0 | 18.19% | 81.81% |

## 3 RESULT AND ANALYSIS

This section dives into the insights gleaned from the selected studies, exploring how scalability principles can inform the successful implementation of a four-day workweek. The analysis is structured and addressing core research objectives:

**Unveiling Workload Scalability:**

The resonance across studies regarding the importance of optimizing resource allocation and workload distribution for seamless transitions is evident. Alfares [2] proposes an optimization algorithm, highlighting potential efficiency gains. Furthermore, [5] shows work redesign for streamlined operations and redefined priorities, echoing cloud computing's scalability principles. Additionally, the concept of dynamic adjustment, akin to cloud resource scaling, emerges as crucial for adapting to varying workloads within a four-day model. Bersin [4] emphasizes asynchronous communication and streamlined operations to manage fluctuating demands effectively. Topp et al. [17] highlight compressed workweek challenges, suggesting flexible and adaptable work structures.

**Examining the Benefits and Drawbacks**

Studies consistently reveal promising benefits associated with a four-day workweek, including increased employee satisfaction, improved work-life balance, and potential productivity boosts [10, 19]. Advanced RV's case study exemplifies these, showcasing sustained positive impacts after implementing a four-day week [9]. However, identified challenges include potential burnout from workload compression, varying impacts across job types, and the need for careful planning and implementation to address individual preferences and work requirements [16]. Studies acknowledge historical challenges with compressed workweeks and the importance of overcoming non-linear work hours-productivity dynamics [10].

**Bridging the Gap: From Challenges to Solutions:**

Pilot programs and clear communication are valuable facilitators. Harvard Business Review [10] recommends resetting employee expectations and providing guidance through communication and coaching. Bersin [4] suggests work redesign, as many young workers challenge traditional norms. Addressing diverse needs and preferences is paramount. Sahadi [16] highlights varying impacts on well-being across job types and individual preferences. Flexibility and adaptability are key elements to consider when designing a four-day workweek model. While these are great solutions, the need to think beyond the norm is important. Just as many organizations have adopted Agile principles. We propose the concept of Workforce scalability by adopting scalability principles from the Cloud and IT.

## 4 DISCUSSION

The exploration of scalability principles inherent in cloud computing and IT has paved the way for envisioning a transformative approach to workforce management for the four-day workweek. Drawing parallels between the dynamic allocation of resources in cloud computing and the need for flexibility in human resource configurations, this discussion synthesizes key features from the literature that shows the potential applicability of scalability concepts to optimize workforce efficiency.

Workforce Scalability: Inspired by cloud resource allocation, this emphasizes dynamic workload distribution and flexible human resource management. Similar to scaling up servers during peak demand, organizations can utilize on-demand talent pools or micro-tasks to address fluctuating workloads within the four-day model. This aligns with [12] suggestion, ensuring optimal resource utilization and preventing burnout risks.

Embracing Agility and Adaptability: Just as cloud systems adapt to diverse needs, workforce agility is crucial for the four-day workweek. Asynchronous communication, championed by [4], and flexible work structures advocated by [17] empower employees to manage their time effectively and contribute meaningfully within compressed work schedules.

Continuous Improvement through Data-Driven Insights: Scalability and Cloud systems rely heavily on data for optimization. They rely on data analytics for optimization, as data-driven insights can inform decisions related to resource allocation, performance tuning, and scalability planning. Similarly, the four-day model necessitates regular feedback mechanisms. Gathering employee input, monitoring productivity metrics, and analyzing workflow efficiency, as suggested by the [10], are crucial for continuous improvement and ensuring the long-term sustainability of this approach.

Elasticity in Workforce Management: Elasticity, a core scalability principle, refers to the ability to scale resources up or down based on industry needs in IT. Translating this to workforce management, organizations adopting a four-day workweek can exhibit elasticity by dynamically adjusting the workforce size, deployment patterns, and competency mix to match evolving demands. This elasticity enables organizations to optimize efficiency while aligning with the scalability principles observed in IT infrastructure.

Addressing Individual Preferences: Scalability in IT involves accommodating user needs and demands [8]. In a similar vein, [16] highlights the importance of addressing individual preferences in the context of a four-day workweek. Acknowledging the varying impacts on well-being across job types, addressing diverse needs, and fostering flexibility are key elements in designing a scalable and adaptable work model.

Lastly, by adopting workforce scalability principles, organizations can unlock the transformative potential of a four-day workweek. This future-proof approach, built on dynamic resource allocation, data-driven insights, and a commitment to individual well-being, paves the way for a more balanced, productive, and ultimately, flourishing work environment.

## 5 CONCLUSION

The proposition of applying scalability principles to foster a four-day workweek represents a paradigm shift in workforce management, aligning with the efficiency gains observed in IT and cloud computing. By optimizing resource allocation, embracing dynamic adjustment, and exhibiting elasticity in workforce management, organizations can navigate the challenges posed by traditional work structures. The literature emphasizes the potential benefits of a four-day workweek, including increased employee satisfaction, improved work-life balance, and enhanced productivity.

The critical success factors include pilot programs for effective communication, work redesign strategies, and an unwavering commitment to addressing diverse needs. As organizations seek to enhance agility, scalability, and overall efficiency, the concept of a four-day workweek emerges as a viable solution. Just as cloud infrastructure dynamically scales to meet demand, organizations can leverage scalability principles to tailor their workforce configurations for optimal performance and employee well-being. By adopting workforce scalability principles, organizations can effectively navigate the transition to a four-day workweek. This framework, built on adaptability, dynamic resource allocation, and data-driven insights, paves the way for a future of work that prioritizes employee well-being, fosters productivity, and unlocks the full potential of a compressed work schedule. The continuous exploration and implementation of such adaptive strategies will be crucial for organizations to thrive in the evolving landscape of modern work structures. The journey towards a flourishing four-day workweek demands a paradigm shift, and workforce scalability serves as a powerful compass, guiding organizations towards a more balanced and productive future.